# Plasma polarization in massive astrophysical objects

**I L Iosilevskiy**
Moscow Institute of Physics and Technology (State University), 141700, Russia
Joint Institute for High-Temperature of RAS, Izhorskaya 13/19, Moscow, 125412, Russia

E-mail: ilios@orc.ru

**Absract.** Macroscopic plasma polarization, which is created by gravitation and other mass-acting (inertial) forces in massive astrophysical objects (MAO) is under discussion. Non-ideality effect due to strong Coulomb interaction of charged particles is introduced into consideration as a new source of such polarization. Simplified situation of totally equilibrium isothermal star without relativistic effects and influence of magnetic field is considered. The study is based on density functional approach combined with "local density approximation". It leads to conditions of constancy for generalized (electro)chemical potentials and/or conditions of equilibrium for the forces acting on each charged specie. New "non-ideality force" appears in this consideration. Hypothetical sequences of gravitational, inertial and non-ideality polarization on thermo- and hydrodynamics of MAO are under discussion.

## 1. Introduction

Long-range nature of Coulomb and gravitational interactions leads to specific manifestation of their joint action in massive astrophysical objects (MAO). The main of them is polarization of plasmas under gravitational attraction of ions. Extraordinary smallness of gravitational field in comparison with electric one (the ratio of gravitational to electric forces for two protons is $\sim 10^{-36}$) leads to the fact that extremely small and thermodynamically (energetically) negligible deviation from electroneutrality can provide thermodynamically noticeable (even significant) consequences at the level of first (thermodynamic) derivatives. This is the main topic of present paper.

## 2. Electrostatics of massive astrophysical objects

Gravitational attraction polarizes plasma of massive astrophysical bodies due to two factors: (*i*) smallness of electronic mass in comparison with ionic one and (*ii*) general non-uniformity of MAO due to long-range nature of gravitational forces. The first, mass-dependent type of gravitational polarization is part of more general phenomenon: (A) - any inertial (mass-acting) force (due to rotation, vibration, inertial expansion and compression *etc*.) polarizes ion-electron plasma due to the same reason: low mass of electron in comparison with that of ions. The second type of discussed polarization is also part of more general phenomenon (see for example [1]): (B) - any non-uniformity in equilibrium Coulomb system is accompanied by its polarization and existence of stationary profile of average electrostatic potential. This potential is thermodynamic quantity because it depends on thermodynamic parameters. Important particular case is existence of stationary drop of average electrostatic potential at any two-phase interface in equilibrium Coulomb system [2]. It is valid for terrestrial applications: two-phase interfaces in ordinary and dusty plasmas, as well as in ionic liquids and molten salts (*Galvani potential*) [3][4]. It is valid also for astrophysical applications: phase boundaries in planets and compact stars [5][6]. It is valid also for simplified Coulomb models [7][8][9]. Equilibrium potential of two-phase interface is thermodynamic quantity: it depends on thermodynamic parameters (bulk properties) of coexisting phases only. In contrast to the electron work function, Galvani potential does not depend on properties of two-phase interface itself: i.e. its form, purity etc. [10].

Remarkable feature of gravitational polarization is that resulting average electrostatic field *must be* of the *same order* as gravitational field (counting per one proton). Average electrostatic force must be equal to *one half* of gravitational force in ideal and non-degenerated isothermal electron-proton plasma of outer layers of a star [11][12]. Average electrostatic force is supposed to be equal just *twice* gravitational force (counting per one proton) in opposite case of ionic plasma on strongly degenerated electronic background in compact stars (white dwarfs, neutron stars etc) [13][5] etc. Ions in thermodynamically equilibrium MAO are suspended, figuratively speaking, in electrostatic field of strongly degenerated and weakly compressible electrons. Exact equality ($F_E^{(p)} = 2F_G^{(p)}$) corresponds to zero order approximation in expansion by small parameter $x_m \equiv m_e/m_i$. This proportionality (*congruence*) of gravitational and average electrostatic fields is not restricted by condition of strong ionization. The same rule is valid for weakly ionized plasmas. The key (dominating) factors for this ratio are (*i*) Coulomb non-ideality and (*ii*) degree of electronic degeneracy.

Real plasmas of compact stars (white dwarfs and neutron stars) are close to isothermal conditions due to high thermal conductivity of degenerated electrons. At the same time plasma of ordinary stars, for example, of the Sun, is not isothermal. Temperature profile, heat transfer and thermo-diffusion exist in such plasmas. It should be taken into account self-consistently in calculation of average electrostatic field.

### 3. Ideal-gas approximation

Plasma polarization at *micro*-level is well known in classical case as Debye-Hueckel screening [15] and in the case of degenerated electrons as Thomas-Fermi screening [16]. Plasma polarization under gravitational forces at *macro*-level is less known although it was claimed [17] and proved at the same years [11][12]. Average electrostatic potential (*Pannekoek - Rosseland electrostatic field*) was calculated for idealized thermodynamically equilibrium, ideal and non-degenerate, isothermal and electroneutral plasma in outer layers of normal stars. Exact relation between electrostatic and gravitational forces was obtained for two-component plasma of electrons and ions of charge $Z$ and atomic number $A$

$$F_E^{(p)} = -F_E^{(e)} = -\frac{A}{(Z+1)} F_G^{(p)} \qquad F_E^{(Z)} = -\frac{Z}{(Z+1)} F_G^{(Z)} \qquad (1)$$

Here $F_E^{(p)}$, $F_G^{(p)}$, $F_E^{(e)}$, $F_E^{(Z)}$, $F_G^{(Z)}$– electrostatic and gravitational forces acting on one proton (*p*), electron (*e*) and ion (*Z*). Ideal-gas formula (1) was extended approximately on conditions of dense non-ideal plasma with highly degenerated electrons in interiors of compact stars by Bilsten *et al.* ([13][14] etc.). New small parameter was introduced into consideration: $x_C$ – ratio of ideal-gas compressibility of electrons and ions at given density. The screening effect proved to be higher in this conditions so that polarization force compensates (screens) almost totally gravitation force acting on each ion at $x_C \ll 1$. For example, average electric force acting on impurity proton is twice higher than gravitation force in conditions of dense mixture of nuclei $\{_{16}O^{8+}, \,_{12}C^{6+}, \,_{4}He^{2+}\}$ on highly degenerate electronic background in interior of typical white dwarfs (WD). It means that such protons are repelled out from degenerate WD-interior until they reach non-degenerate conditions near the surface (for example [14]).

$$F_E^{(p)} = -F_E^{(e)} \cong -\frac{A}{Z} F_G^{(p)} \qquad F_E^{(Z)} \cong -F_G^{(Z)} \qquad (2)$$

It seems natural to suppose that in general case the value of discussed compensation lay between two limits (1) and (2). The point of present paper is that it is valid for ideal-gas assumption only (with arbitrary degree of electron degeneracy). In fact, it is not correct if one takes into account non-ideality effects. There may be conditions when polarization force *overcompensates* gravitational force due to additional non-ideality effect, i.e. $|F_E^{(Z)}| \gtrsim |F_G^{(Z)}|$ (see below).

### Gravitational polarization with non-ideality effects

Our goal is to improve presently existing approach to describe gravitational polarization in strongly non-ideal plasma, which is typical for planets and compact star's interior. Approach accepted by [13][5][18] etc. operates with idea of individual *partial pressures* for ions and electrons, $P_i(n_i,T)$ and $P_e(n_e,T)$, and based on solution of several separate ("*partial*") *hydrostatic equilibrium equations* for each specie of particles instead of *unique* hydrostatic equilibrium equation for *total* pressure and total mass density in standard approach [19][20]. The point of present work is that partial pressures and partial hydrostatic equilibrium equations are not well-defined quantities in general case of equilibrium non-ideal system.

Let's consider simplified case of hypothetical non-uniform self-gravitating body in total thermodynamic equilibrium without relativistic effects and influence of magnetic field. General approach for description of thermodynamic equilibrium in this case is multi-component variational formulation of statistical mechanics [21][22][23]. Thermodynamic equilibrium conditions may be written in three forms: (*i*) – extremum condition for thermodynamic potential of total system (free energy functional) regarding to variations of one-, two-, three-particle etc. correlations in the system; (*ii*) – constancy conditions for generalized "electro-chemical" potentials [2] for all species (electrons, ions etc.), and (*iii*) – zero conditions for the sum of (generalized) average forces acting on each specie of particles in the system. The problem is that all three values: (total) thermodynamic potential, (partial) electrochemical potentials and (partial) average forces are essentially non-local functionals on mean-particle correlations. Next (standard) technique is separation of main non-local parts of free energy functional – electrostatic and gravitational energies in mean-field approximation.

$$F\{T,V,N_j,N_k,...\} = \min F\left(T,V\,|\,\left[\{n_i(\cdot)\},\{n_{ij}(\cdot,\cdot)\}\right]\right) \equiv$$
$$\equiv \min\left(-\sum_{j,k}\frac{Gm_jm_k}{2}\int\frac{n_j(\mathbf{x})\cdot n_k(\mathbf{y})}{|\mathbf{x}-\mathbf{y}|}d\mathbf{x}d\mathbf{y} + \sum_{j,k}\frac{Z_jZ_ke^2}{2}\int\frac{n_j(\mathbf{x})\cdot n_k(\mathbf{y})}{|\mathbf{x}-\mathbf{y}|}d\mathbf{x}d\mathbf{y} + F^*\left[T,V\,|\,\{n_j(\cdot)\},\{n_{jk}(\cdot,\cdot)\}\right]\right) \quad (3)$$

Here $n(\mathbf{x})$ and $n(\mathbf{x},\mathbf{y})$ – one- and two-particle densities. It is assumed that all non-local effects are exhausted by first two terms of right-hand side of (3). Consequently next widely used technique is the "local-density" approximation (4) for free energy term $F^*[...]$ of hypothetical non-ideal charge system with extracted electrostatic and gravitational energies in mean-field approximation.

$$F^*\left[T,V\,|\,\{n(\cdot)\}\right] \approx \int_V f^*\left(n_j(\mathbf{r}),n_k(\mathbf{r}),...,T\right)d\mathbf{r} \quad,\quad f^*\left(n_j,n_k,...,T\right) \equiv \lim\left\{\frac{F^*\left(N_j,N_k,...,V,T\right)}{V}\right\}_{\{N_j,N_k,...\}\to\infty,V\to\infty}^{(N_j/V)\to n_j,(N_k/V)\to n_k,...} \quad (4)$$

It should be stressed [24][25] that (local) free energy density $f^*(n_i,n_k,...,T)$ in (4) must be defined as thermodynamic limit of specific free energy of (new) *uniform* macroscopic *non-electroneutral* multi-component charge system with charge particle densities $(n_j, n_k,...)$ *on compensating Coulomb (and strictly speaking gravitational) background(s)* [1][26]. It means that we deal with free energy $F^*(\mathbf{N}, T)$ of artificial system with additional attraction, which could be, formally speaking, thermodynamically unstable in conditions of strong Coulomb non-ideality [1], i.e. matrix $||\partial^2 F^*/\partial \mathbf{N}^2|| \equiv ||\partial \boldsymbol{\mu}^{(chem)}/\partial \mathbf{n}||$ could lose its positiveness in strong non-ideality conditions ($\Gamma \gg 1$, see below). It should be stressed that it does not mean thermodynamic instability of *whole* non-uniform Coulomb system (star), which is stabilized in long-wave limit via mean-field Coulomb term in (3). Nevertheless artificial short-wave instability is still remain in equilibrium ionic profile $n_i(\mathbf{r})$ due to local density approximation (4). It should be suppressed by addition of corresponding gradient terms in free energy density functional $F^*$ in (3). The discussed artificial short-wave instability could be avoided also in frames of local density approximation (4) via special choice of long-range potential in the mean-field Coulomb term in (3) [27] Both these tricks are not necessary in context of present paper for general illustration of non-ideality influence on plasma polarization in high-gravity astrophysical objects.

Thermodynamic equilibrium condition in integral form (4) leads to two sets of corresponding local forms in terms of electrochemical potentials (5) and in terms of generalized thermodynamic forces (6):

$$m_j\varphi_G(\mathbf{r}) + q_j\varphi_E(\mathbf{r}) + \mu_j^{(chem)}\{n_i(\mathbf{r}), n_e(\mathbf{r}); T\} = \mu_j^{(el.chem)} = \text{const} \qquad (j = \text{electrons, ions}) \quad (5)$$

$$m_j\nabla\varphi_G(\mathbf{r}) + q_j\nabla\varphi_E(\mathbf{r}) + \nabla\mu_j^{(chem)}\{n_i(\mathbf{r}), n_e(\mathbf{r}); T\} = \nabla\mu_j^{(el.chem)} = 0 \qquad (j = \text{electrons, ions}) \quad (6)$$

Here $\varphi_G(\mathbf{r})$ и $\varphi_E(\mathbf{r})$ – gravitational and electrostatic potentials, $\mu_j^{(chem)}$ and $\mu_j^{(el.chem)}$– are the local chemical and non-local electrochemical potentials, $m_j$ and $q_j$ – mass and charge of specie $j$ ($j = i,e$), $\nabla\varphi(\mathbf{r})$, $\nabla\mu(\mathbf{r})$ – spatial gradients. It should be stressed that the set of equations (5) and/or (6) are well-defined equivalents (substitute) for the set of separate equations of hydrostatic equilibrium for mentioned above partial pressures and densities of charged species in ideal-gas conditions [11][12].

Thermodynamic equilibrium conditions in form (6) with electroneutrality condition lead to final equation for average electrostatic field in simplified case of two-component non-ideal electron-ionic plasma with arbitrary degree of non-ideality and electron degeneracy [26].

$$Ze\nabla\varphi_E(\mathbf{r})\left[1+\frac{\Theta}{Z}\right] = -M\nabla\varphi_G(\mathbf{r})\left[1-m\frac{\Theta}{M}\right], \quad \Theta \equiv \frac{(\mu_{ii}^0 + \Delta_i^i + Z_i\Delta_e^i)}{(Z\mu_{ee}^0 + Z\Delta_e^e + \Delta_i^i)}, \quad \Delta_k^j \equiv \left(\frac{\partial\Delta\mu_j^{chem}}{\partial n_k}\right)_{T,n_{i\neq k}}, \quad \mu_{jj}^0 \equiv \left(\frac{\partial\mu_j^0}{\partial n_j}\right)_{T,n_{k\neq j}} \quad (7)$$

Here $m, M, Z$ – masses and charge of electrons and ions. $\mu_j^0(n_j,T)$ and $\Delta\mu_j^{(chem)}(n_i,n_e,T)$ – ideal and non-ideal part of *local* chemical potential of specie $j$ ($j = i,e$). Note that for Coulomb interaction non-ideal corrections $\Delta\mu^{(chem)}$ and their derivatives $\Delta_k^j$ in (7) are negative.

*Comments. Thermodynamics:*

- In ideal-gas approximation equation (7) reproduces both known limits, (1) and (2), with non-degenerated or highly-degenerated electrons correspondingly, and gives *monotonic* growth of screening effect from (1) to (2) at intermediate degree of electronic degeneracy.

$$F_E^{(Z)} = -\frac{(1 - x_c m M^{-1})}{(1 + x_c Z^{-1})} F_G^{(Z)} \approx -\left(1 + \frac{x_c}{Z}\right)^{-1} F_G^{(Z)}, \qquad x_c(\zeta_e) \equiv \left(\frac{\mu_{ii}^0}{Z \mu_{ee}^0}\right), \qquad \zeta_e \equiv n_e \lambdabar_e^3, \qquad (7i)$$

($1 \geq x_c \geq 0$ when $0 \leq \zeta_e \leq \infty$). Simplified approximation for $x_c(\zeta_e)$ could be used for estimations:

$x_c(\zeta_e) \approx (1 + a_1 \zeta + a_2 \zeta^2)^{2/3} (1 + b_1 \zeta + b_2 \zeta^2)^{-1}$ (here $a_1, a_2, b_1, b_2 \approx$ 0.21314, 0.02827, 0.28418, 0.04712)

In ultrahigh densities ($\rho \geq 10^6$ g/cc), when electronic subsystem became relativistic [19] [20], useful approximation for $x_c(\zeta_e)$ could be found elsewhere [28].

- In contrast to the ideal-gas approximation (7i) equation (7) in general case describes equilibrium conditions as competition between not *two,* but *three* sources of influence: gravitation field, polarization field and generalized "*non-ideality force*".
- Coulomb "non-ideality force", when it is taken into account in (7), moves positive ions *inside* the star in addition to gravitation. Hence "non-ideality force" *increases* compensating electrostatic field $\Delta\varphi_E(\mathbf{r})$ in comparison with ideal-gas approximation (7i).
- In the case of classical (non-degenerated) plasma the function $\Theta$ in (7) depends on Coulomb non-ideality. In the weak non-ideality limit for two-component electron-ionic($Z$) plasma it could be described in Debye-Hückel approximation

$$F_G^{(Z)} \approx -F_E^{(Z)} \left[1 + \frac{(1 - Z^2 \Gamma_D / 4)}{Z(1 - \Gamma_D / 4)}\right], \quad \Gamma_D \equiv (e^2 / kTr_D) \ll 1, \quad \zeta_e \equiv n_e \lambdabar_e^3 \ll 1, \quad \{r_D^{-2} \equiv (4\pi e^2(1 + Z^2)/kT)\} \quad (7ii)$$

- It is non-symmetry in thermodynamic properties of electrons and ions, who manifests itself in discussing "non-ideality force". For example, in *symmetrical* classical (non-degenerated) electron-protonic system Coulomb non-ideality corrections in numerator and denominator of non-ideality function $\Theta$ in (7) cancel each other totally, so that $\Theta = 1$ and resulting polarization field is equal to its ideal-gas limit (1) at *any* degree of Coulomb non-ideality. [$F_E^{(p)} = -\frac{1}{2} F_G^{(p)}$]
- The non-ideality function $\Theta$ may be negative and the bracket term $[1 + (\Theta/Z)]$ in right side of (7) may be *less* than *unity* in the case when strongly non-ideal ionic subsystem is combined with highly degenerated and almost ideal electrons (for example, in white dwarfs). In this case one meets "*overcompensation*" when polarization field could be *higher* (by absolute value) than gravitation field: i.e. $|F_E^{(Z)}| > |F_G^{(Z)}|$. Rough approximation (7iii) could be useful. More accurate approximation for EOS of OCP could be found elsewhere (see for example [29]).

$$F_G^{(Z)} \approx -F_E^{(Z)} \left[1 - \frac{a_M \Gamma_Z}{Z} x_c(\zeta_e)\right], \quad \{\Gamma_Z \equiv Z^2 e^2 (4\pi n_i / 3)^{1/3} / kT \gg 1, \ \zeta_e \equiv n_e \lambdabar_e^3 \gg 1, a_M \approx 0.4\} \quad (7iii)$$

- Any jump-like discontinuity in local thermodynamic state, in particular phase transition interface or the set of interfaces between mono-ionic layers with different $M_i$, $Z_i$ and $\Gamma_Z$ in neutron star crust [30], leads in general case to corresponding jump-like discontinuity in "non-ideality force" in (7) and consequently, to jump-like discontinuity in final polarization field $\Delta\varphi_E(\mathbf{r})$. It means in its turn appearance of *macroscopic charge* at all discussed mean-phase and mean-layer *interfaces* in addition to electrostatic potential drop (Galvani potential) mentioned at the beginning of the paper.
- Equation (7), which connect polarization field with gravitation and non-ideality forces, could be generalized for the case of ionic mixture ($Z_1, Z_2, \ldots$) [27]. In this case average polarization field repels out ions with smaller ratio *A/Z* and pulls inside the ions with higher ratio *A/Z*. In addition polarization field pulls inside the ions with higher charge $Z$ due to non-ideality effects. For example, such repulsion of minor proton impurity from fully ionized helium plasma in outer layer of white dwarfs and neutron stars prevents proton diffusion and subsequent burning in deeper layers of a star (see for example [14]).
- Equation (7) is not restricted by spherical symmetry conditions. It is valid for rotating stars and stars in binary systems *etc*. It is valid for any self-gravitating system in *total thermodynamic equilibrium* (see above comment about non-isothermal state)

- Strong correlation of gravitation and polarization fields (7) is not restricted by condition of high degree of plasma ionization. Weakly ionized but non-ideal plasmas of outer layers of self-gravitating bodies (planets, stars etc.) must obey the same equations (3-7). Multi-component variant of equations (4-7) should be used (so-called "chemical picture") in this case and Eq (7) will include contributions from neutral-charge interactions, i.e. not just the Coulomb non-ideal terms.

*Comments. Hydrodynamics:*
- Plasma polarization in MAO leads to noticeable hydrodynamic consequences.
- Plasma polarization could suppress hydrodynamic instabilities in MAO. For example, plasma polarization could suppress hypothetical Rayleigh-Taylor instability in liquid mixture of nuclei $\{_{16}O^{8+}, {}_{12}C^{6+}, {}_{4}He^{2+}\}$ in interior of typical white dwarf in the vicinity of its freezing boundary [31]. Accordingly (7) polarization field compensates almost totally gravitation field acting on any nucleus, O, C and He, due to their symmetry ($A/Z = 2$) so that the total force is roughly equal to zero: $(F_E^{(Z)} + F_G^{(Z)} \approx 0)$. The final weak discrimination in total force acting on each ion in the mixture $\{_{16}O^{8+} + {}_{12}C^{6+} + {}_{4}He^{2+}\}$ depends on interplay between Coulomb non-ideality effects and electron degeneracy.
- Besides gravitation any inertial (mass-acting) force (rotation, vibration, inertial expansion (explosion) and compression (collapse) *etc*.) could polarize electron-ionic plasma due to the same reason – low mass ratio of electrons and ions.
- Effect of rotation of MAO (including differential one) could be taken into account naturally in discussing form of equilibrium conditions (3)-(7). For this purpose rotation energy functional, local centrifugal potential and local centrifugal force should be added to the total free energy functional (3) and should be included in electrochemical potential (5) and dynamic equilibrium equation (6) correspondingly. For example, polarization field should be equal to zero in the case of rotation limit when centrifugal force is equal to gravitational one.
- Any acoustic oscillation slow enough relatively to electronic and ionic relaxation must be accompanied with electron-ionic polarization and consequent electromagnetic oscillation. Hence acoustic properties of a star must include in general case dependence on parameters of Coulomb interaction: ionic charge $Z$, ratio $A/Z$, Coulomb non-ideality parameter etc.

*Comments. Polarization parameters:*
Proportionality (congruence) of average electrostatic and gravitational fields in a star means that excess charge profile, while being very small $[Q(\mathbf{r}) <<< \rho(\mathbf{r})]$ is proportional (congruent) to the density profile of the star $[(Q(\mathbf{r}) \sim \rho(\mathbf{r})]$. It gives simple estimation of contour parameters of discussed polarization (Table 1): maximal value of electrostatic field $E_{max}$ (at the surface) and maximal value of electrostatic potential $U_{max}$ (in the center).

$E_{max}(r = R) \cong gm_p/e = (GMm_p/R^2e) \approx 2.85 \cdot 10^{-8} [M^*/(R^*)2]$ V/cm

$U_{max}(r = 0) \cong gm_pR/2 = (GMm_p/2R) \approx 1 \cdot 10^3 (M^*/R^*)$ eV

Here $M^* \equiv M/M_\odot$; $R^* \equiv R/R_\odot$ ($M_\odot \cong 1.99 \cdot 10^{33}$ g, $R_\odot \cong 6.96 \cdot 10^{10}$ cm - mass and radius of the Sun)

**Table 1.** Estimated parameters of average electrostatic potential of stars

|  | Sun | White Dwarf | Neutron Star |
|---|---|---|---|
|  | $M \equiv M_\odot, R \equiv R_\odot$ | $M_{WD} = M_\odot, R_{WD} = R_{Earth}$ | $M_{NS} = M_\odot, R_{NS} = 10$ km |
| $U_{max}$ [eV] | 1 keV | 1 MeV | 70 MeV |
| $E_{max}$ [V/cm] | $3 \cdot 10^{-8}$ | 0.03 | 150 |

One can conclude that contour parameters of discussed polarization are extremely small for non-exotic MAO (stars, planets etc.) [32]. At the same time they are noticeable and even significant in exotic situation in neutron and combine (*strange*) stars [5][18], combine (strange) white dwarfs [33] etc.

**Acknowledgments**
The work was supported by Grants: CRDF № MO-011-0, ISTC 3755 and by RAS Scientific Programs "Research of matter under extreme conditions" and "Physics of high pressure and interiors of planets"